\documentclass[aps,prl,reprint,superscriptaddress]{revtex4-1}

\usepackage{graphicx}
\usepackage{bm}
\usepackage{multirow}
\usepackage{sidecap}
\usepackage{wrapfig}
\usepackage{xcolor}

\usepackage{amsmath}
\usepackage{amssymb}
\usepackage{upgreek}

\usepackage{natbib}
\begin{document}

\title{Sum-frequency generation through a unique Feynman diagram formalism:\\ the case of bipartite organic/inorganic complexes}


\author{T. Noblet}

\affiliation{University of Paris-Sud, Universit\'e Paris-Saclay, Laboratoire de Chimie Physique, CNRS, B\^atiment 201P2, 91405 Orsay, France.}

\author{C. Humbert}

\affiliation{University of Paris-Sud, Universit\'e Paris-Saclay, Laboratoire de Chimie Physique, CNRS, B\^atiment 201P2, 91405 Orsay, France.}

\date{\today}

\begin{abstract}
In quantum electrodynamics, optical processes are theoretically described by double-sided Feynman diagrams. This formalism is powerful in the case of molecules but proves inappropriate to account for light-matter interactions within complex hybrid systems constituted of organic and inorganic matter. The double-sided Feynman diagrams do not easily enable to implement the coupling between the electronic properties of the former and the vibrational response of the latter. Here we present a new general method bridging optics and condensed matter physics in order to properly account for the underlying fundamental process thanks to the classical Feynman diagrams dedicated to solid state physics, instead of the double-sided diagrams commonly used in nonlinear optics. In a manner both rigorous and pedagogical, we especially show how Feynman diagrams can be used to analytically derive the quantum expressions of second-order optical response functions which prove to be in complete agreement with previously established experimental results.
\end{abstract}


\maketitle


Since the emergence of light quantum theory \cite{Glauber2006} and for several decades now, nonlinear optics has unveiled the existence of fine couplings within matter \cite{Bloembergen1982,Franken1963} thanks to many laser-based techniques of optical characterization, such as second harmonic generation \cite{Franken1961,Chen1973}, sum-frequency generation (SFG) \cite{Guyot1987,Shen1989,Bonn2000,Stiop2011} and Raman scattering \cite{Glocker1943}. In the latter two, optics enables to probe and evidence couplings between the electronic and vibrational structures of systems of interest. So far, these have been extensively studied for molecular systems \cite{Xu1999,Zheng2006,Guthmuller2008}, in which the vibronic structure can be directly examined by Raman and SFG spectroscopies \cite{Raschke2002,Hayashi2002,Dreesen2004bis,Hung2015}. In the point of view of quantum mechanics, such optical processes are described and implemented on the base of double-sided Feynman diagrams \cite{Shen,Boyd,Ward1965}. However, within hybrid systems made of organic and inorganic building blocks \cite{Noblet2018}, the description of the vibro-electronic coupling processes suffers from a lack of continuity from the quantum formalism of nonlinear optics (double-sided Feynman diagrams) to that of solid state physics (classical Feynman diagrams \cite{Mahan}). Here we develop a global method based on classical Feynman diagrams in order to address this issue by connecting optics and solid state physics. To validate the accuracy of this theoretical approach, we first check its compatibility with linear optics (accounting for absorption and emission) prior to applying it to nonlinear optics in the light of experimental SFG results recently obtained with a complex hybrid system made of ligand-conjugated quantum dots (QDs) \cite{Noblet2018}. Given the generality of the present work, it is worth noting that the specific case of QDs constitutes an illustration of the method, which is applicable in this way to other hybrid systems, provided the eigenstates of the subsystems and the hamiltonian describing their interactions are known. The method can obviously be employed to test some guesses of the interaction hamiltonian by comparison of the theoretical prediction with the experimental results obtained through SFG spectroscopy. In our case, the assumption of a dipolar coupling between QDs and ligands leads to the analytical computation of a second-order susceptibility compatible with the experiment.

\begin{figure*}
\includegraphics[width=8cm]{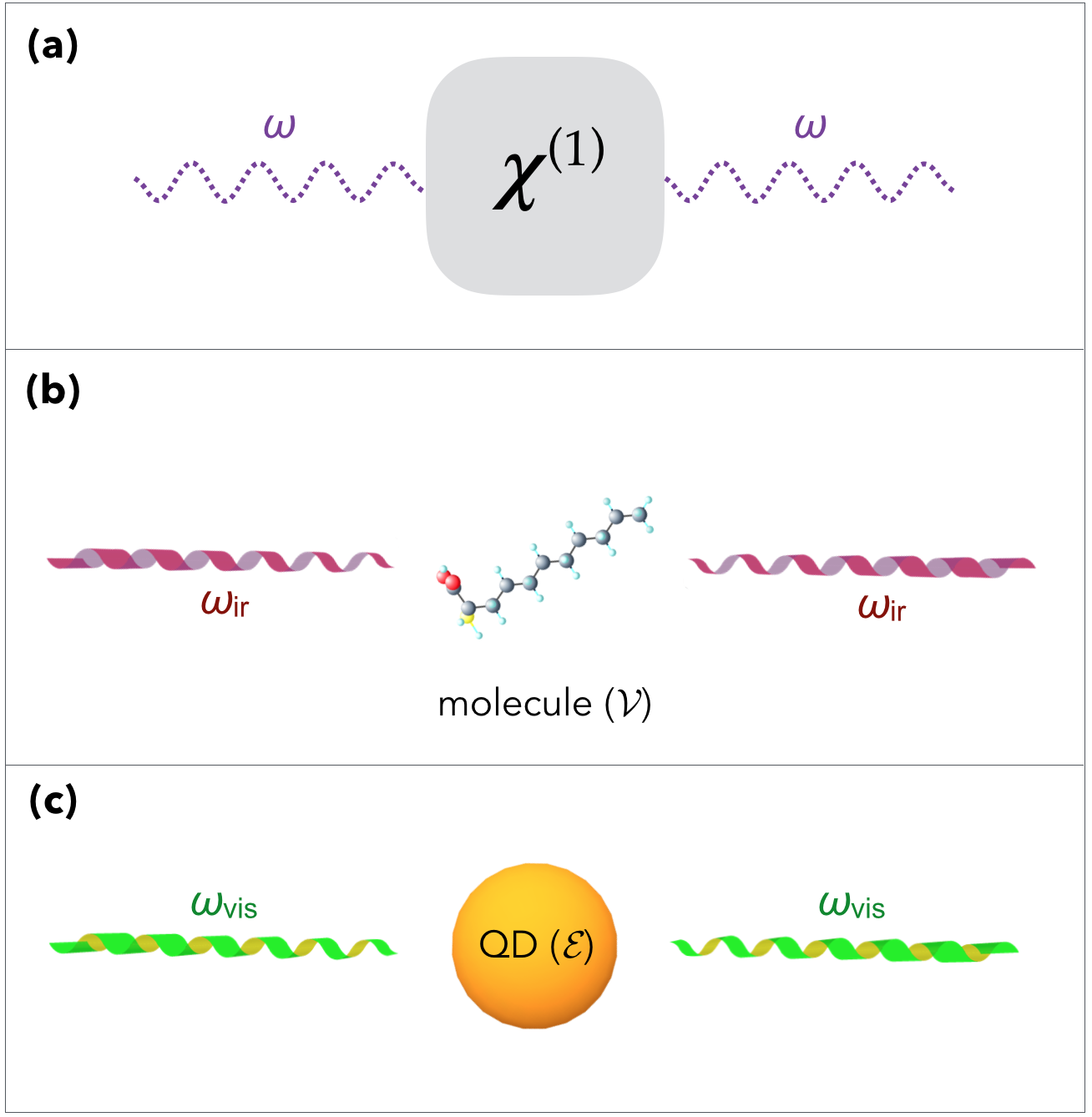} \hskip5mm \includegraphics[width=8cm]{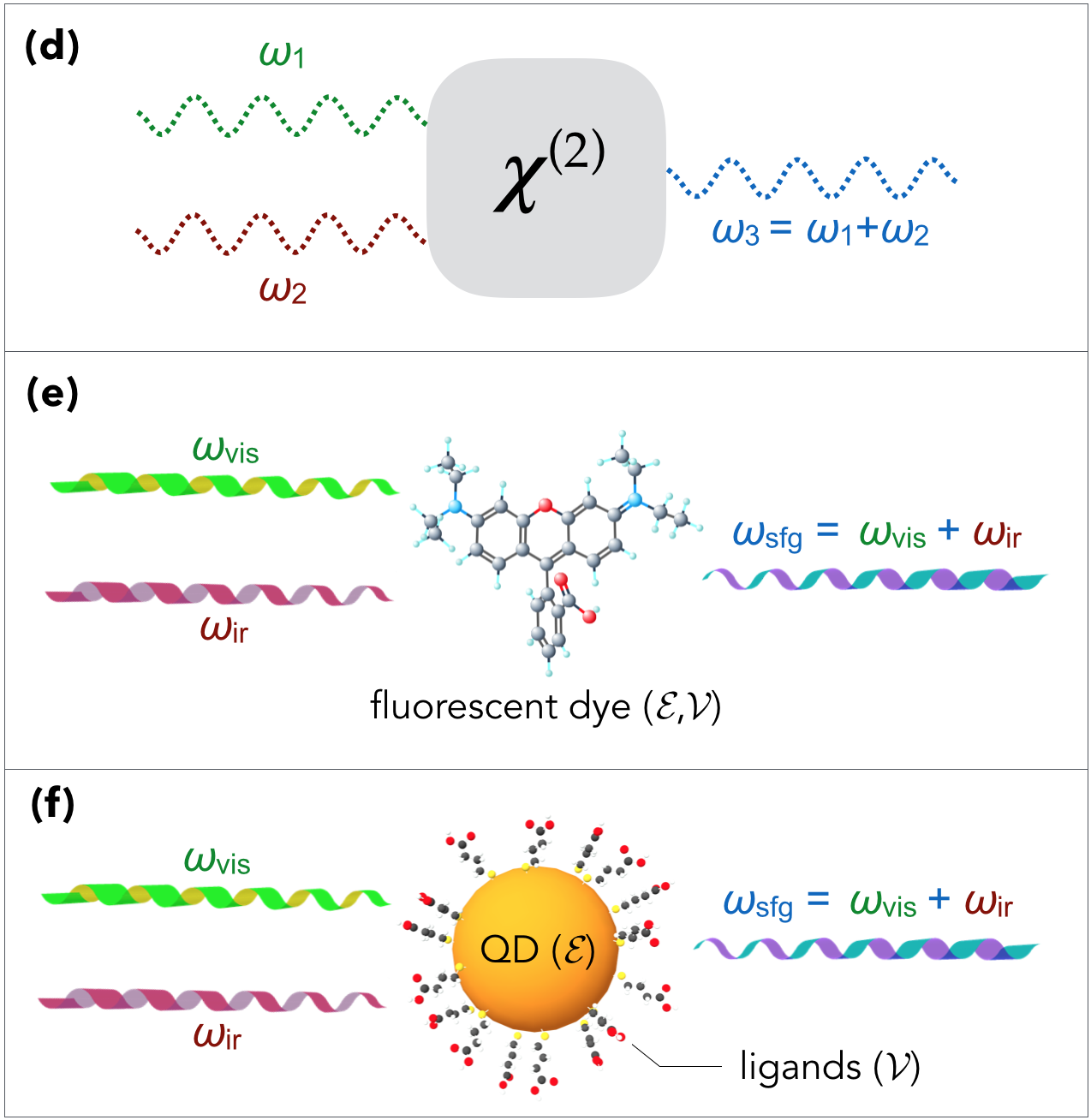}
\caption{\textbf{\textsf{Pictures of linear and nonlinear optics. a}}, The first-order susceptibility corresponds to the Green's function which measures the coherence of a photon $\omega$ along its propagation through a given material. \textbf{\textsf{b}}, Principle of infrared spectroscopy performed on an organic system. Its response function $\bm{\chi}^{(1)}(\omega_\text{ir})$ can be derived from the Green's functions associated to its vibrational states $(\mathcal{V})$. \textbf{\textsf{c}}, Principle of UV-visible spectroscopy performed on an inorganic system. Its response function $\bm{\chi}^{(1)}(\omega_\text{vis})$ can be derived from the Green's functions associated to its electronic states $(\mathcal{E})$. \textbf{\textsf{d}}, The second-order susceptibility corresponds to the Green's function which combines the two input frequencies, $\omega_1$ and $\omega_2$, into $\omega_3$. \textbf{\textsf{e}}, Principle of SFG spectroscopy performed on a molecular fluorescent dye. It consists in mixing a visible and an infrared beams, and measuring the optical signal generated at the sum-frequency. In this conventional case, the two subsystems $(\mathcal{E})$ and $(\mathcal{V})$ are part of the same molecule. \textbf{\textsf{f}}, Principle of SFG spectroscopy performed on an organic/inorganic hybrid system, that is ligand-capped CdTe QDs for instance. In this unconventional case, the visible beam probes the electronic structure of QDs, which constitutes subsystem $(\mathcal{E})$, while the infrared one probes the vibrational structure of ligands, which constitutes subsystem $(\mathcal{V})$.}\label{principle}
\end{figure*}


\begin{figure*}
\includegraphics[width=14cm]{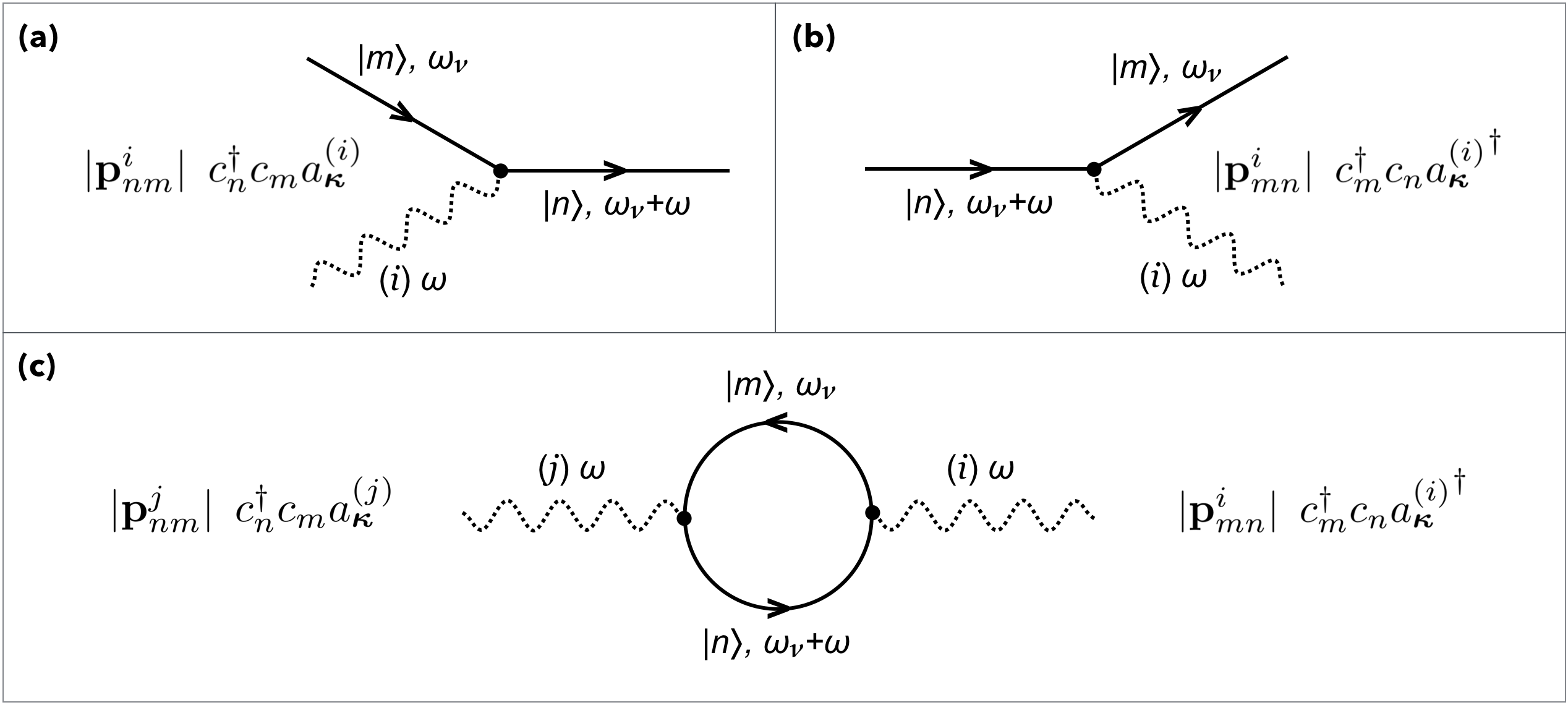}
\caption{\textbf{\textsf{Light-matter interaction and linear response. a}}, Diagrammatic representation of the absorption of an $i$-polarized photon of Matsubara frequency $\imath\omega$ (Methods, equation (\ref{abs})). \textbf{\textsf{b}}, Diagrammatic representation of the emission of an $i$-polarized photon of Matsubara frequency $\imath\omega$ (Methods, equation (\ref{abs})). \textbf{\textsf{c}}, Diagrammatic representation of $\chi^{(1)}_{ij}(\imath\omega)$ for a given quantum system (QD, molecule, atom) of dipole moment $\mathbf{p}$. It is obtained thanks to the combination of the vertices given in Fig. \ref{vertex1}a \& b.}\label{vertex1}
\end{figure*}

\vskip7mm

\noindent\textbf{\textsf{Results}}

\noindent \textbf{Second quantization of simple and hybrid systems.} Within the formalism introduced in Methods, the linear and nonlinear optical response functions $\chi_{ij}^{(1)}$ and $\chi_{ijk}^{(2)}$ of any simple or complex systems are expanded on the base of imaginary time Green's functions. To show the interest and the effectiveness of the approach, we consider in this article the cases of organic (\emph{e.g.} molecules, Figure \ref{principle}b \& e), inorganic (\emph{e.g.} quantum dots, Figure \ref{principle}c) and hybrid systems (\emph{e.g.} ligand-conjugated quantum dots, Figure \ref{principle}f). In Figure \ref{principle} and the following, the notations $(\mathcal{E})$ and $(\mathcal{V})$ depict any set of electronic and vibrational states, respectively. In all generality, we thus describe organic/inorganic hybrid systems as made of an electronic subsystem $(\mathcal{E})$ and a vibrational subsystem $(\mathcal{V})$.  Hence, to be relevant, the method first requires to derive the proper quantum description of the two interacting subsystems $(\mathcal{E})$ and $(\mathcal{V})$. In order to present a universal description, we consider here canonical hamiltonians in the framework of the second quantization formalism.

Each quantum system is characterized by its eigenenergies $\epsilon_n$, whether its states are electronic or vibrational. By introducing the associated creation and annihilation operators, $c_n^\dagger$ and $c_n$, the hamiltonian takes the standard form:
\begin{equation}\label{eqG}
\mathcal{H} = \sum_n \epsilon_n\ c_n^\dagger c_n.
\end{equation}
For instance, the excitonic hamiltonian of QDs can be written as:
\begin{equation}\label{eqE}
\mathcal{H}_\text{QD} = \sum_{p\in (\mathcal{E})} \epsilon_{p}^x\ x_{p}^\dagger x_p,
\end{equation}
where $x_{p}^\dagger$ and $x_{p}$ respectively creates and annihilates an exciton of energy \cite{Haug}:
\begin{equation}
\epsilon_p^x = E_g + \frac{(\hbar\alpha_p^e)^2}{2m_eR^2} +\frac{(\hbar\alpha_p^h)^2}{2m_hR^2}.
\end{equation}
For a spherical QD of radius $R$, $\alpha_p^{e/h}$ depicts the zeros of the spherical Bessel functions associated to the wave functions of the electron, $e$, and the hole, $h$, which compose the exciton \cite{Haug}. $m_e$ and $m_h$ correspond to their respective effective masses. In the following, we assume that $p$ is ordered by increasing energy, \emph{i.e.} $\epsilon_{p}^x< \epsilon_{p+1}^x$, and that $p=0$ depicts the ground state, \emph{i.e.} vacuum of exciton, with zero energy.

For molecules, we make the choice to itemize the quantum states, both electronic and vibrational, by the integer $\sigma$, and to describe them by the fermion operators $d_\sigma^\dagger$ and $d_\sigma$:
\begin{equation}\label{eqV}
\mathcal{H}_\text{mol.} = \sum_{\sigma\in(\mathcal{E},\mathcal{V})} \epsilon_\sigma\ d^\dagger_\sigma d_\sigma.
\end{equation}
Here again, we assume that the eigenenergies $\epsilon_\sigma$ are known, precondition required for identifying the imaginary Green's functions.

Given their generality, equations (\ref{eqE}) and (\ref{eqV}) do not only describe the case of QD/ligand hybrid systems but can be used to account for all subsystems $(\mathcal{E})$ and $(\mathcal{V})$.

\vskip3mm
\noindent\textbf{Derivation of linear response functions.} For any simple system described by equation (\ref{eqG}), we consider the Feynman diagram drawn in Figure \ref{vertex1}c. In virtue of the second quantization of light-matter interactions reminded in Methods, this diagram is built with the vertices associated to the absorption of a $j$-polarized photon of Matsubara frequency $\imath\omega$ and the emission of an $i$-polarized photon of Matsubara frequency $\imath\omega$. As evidenced by the following derivation, this picture gives access to the linear susceptibility $\bm{\chi}^{(1)}$ of a given system of dipole moment $\mathbf{p}$. Indeed, from Feynman rules \cite{Zagoskin,Mahan}, the diagram is converted into:
\begin{equation}\label{x1}
\chi_{ij}^{(1)}(\imath\omega) = - \frac{1}{\beta} \sum_{n,m,\nu} |\mathbf{p}_{mn}^i| |\mathbf{p}_{nm}^j|\ \mathfrak{G}_{mm}(\imath\omega_\nu)\mathfrak{G}_{nn}(\imath\omega_\nu+\imath\omega).
\end{equation}
From equation (\ref{matsuGreen's}), in Methods, and thanks to Matsubara's theorem (see Supplementary Note 1):
\begin{equation}
\chi_{ij}^{(1)}(\imath\omega) = - \sum_{n,m}\ |\mathbf{p}_{mn}^i| |\mathbf{p}_{nm}^j|\ \frac{\rho(\omega_m)-\rho(\omega_n-\imath\omega)}{\omega_m-\omega_n+\imath\omega}. 
\end{equation}
As $\imath\omega$ is the Matsubara frequency of a photon, that is a boson: $\rho(\omega_n-\imath\omega)=\rho(\omega_n)$. Assuming that the system is in its ground state $|0\rangle$ at equilibrium, \emph{i.e.} $\rho(\omega_n)=\delta_{n 0}$, and setting the ground state energy to zero:
\begin{equation}\label{chi1}
\chi_{ij}^{(1)}(\omega) = -\sum_n \left(\frac{|\mathbf{p}_{0n}^i||\mathbf{p}_{n0}^j|}{\omega-\omega_n +\imath0^+} - \frac{|\mathbf{p}_{0n}^i||\mathbf{p}_{n0}^j|}{\omega+\omega_n +\imath0^+} \right).
\end{equation}
According to the nature of the system, $n$ describes electronic and/or vibrational states. Thus, we retrieve the common formula of $\chi_{ij}^{(1)}$ obtained within the density matrix formalism \cite{Shen} and the time-dependent perturbation theory \cite{Haug}.

The convergence between our method based on imaginary time Green's functions and the common approaches that we have just mentioned indicates that our initiative is at least as appropriate as the techniques based on double-sided diagrams. The relevance of the method in the first case of linear optics strongly suggests it is likely to give rise to a brand new way of thinking and describing nonlinear optics. It especially proves that the condensed matter picture of Feynman diagrams is well suited for performing analytical computations of optical response functions, and that imaginary time Green's functions are not only dedicated to solid state physics: they are of great interest for optics too. Consequently, this observation naturally leads us to examine the case of higher-order response functions. As a matter of fact, our method releases all its effectiveness for nonlinear optics.

\vskip3mm
\noindent\textbf{Nonlinear sum-frequency generation.} Probing matter with intense light allows studying its behaviour beyond the linear dielectric regime. One of the highest stakes then lies in the theoretical understanding of the nonlinear response of organic/inorganic hybrid systems. To date, there is not any analytical description, whereas the experiment evidences the existence of vibroelectronic couplings within these systems, thanks to SFG spectroscopy \cite{Humbert2015,Sengupta2017,Noblet2018}. Here, we show that our Feynman diagram-based method takes up the challenge and enables to compute the associated second-order susceptibility $\bm{\chi}^{(2)}=(\chi_{ijk}^{(2)})$. In particular, we choose to illustrate our approach with the case of ligand-capped CdTe QDs \cite{Noblet2018}. As a second-order response function, each $\chi_{ijk}^{(2)}$ is a Green's function which combines two frequencies, $\omega_1$ and $\omega_2$ \cite{Shen, Boyd}. We thus need to draw a Feynman diagram which represents how the hybrid system mixes $\omega_1$ and $\omega_2$ to generate $\omega_3=\omega_1+\omega_2$. Figure \ref{principle}d illustrates the SFG process. The aim of the paper is then to build a Feynman diagram, based on the imaginary time Green's functions associated to the two subsystems $(\mathcal{E})$, from the QDs, and $(\mathcal{V})$, from the ligands, in order to relate $\omega_1$, $\omega_2$ and $\omega_3$. The complete Green's function computed from this new diagram will be therefore interpreted as $\chi_{ijk}^{(2)}(\omega_1,\omega_2)$.

In this context, Figures \ref{principle}e \& f explicitly show the difference of point of view between (e) purely molecular and (f) hybrid systems. In the first case, the electronic and the vibrational states are coupled within the same molecule, and the double-sided Feynman diagrams efficiently allow the calculation of $\bm{\chi}^{(2)}(\omega_1,\omega_2)$ \cite{Huang1994}. In the second case, $(\mathcal{E})$ is associated to the inorganic subsystem, while $(\mathcal{V})$ corresponds to the organic one. In the following, we dwell on this unconventional hybrid configuration, but a complete derivation of $\bm{\chi}^{(2)}(\omega_1,\omega_2)$ in the conventional case of molecular systems is achieved in Supplementary Note 2, thanks to our diagrammatic method. 



\begin{figure}
\includegraphics[width=7cm]{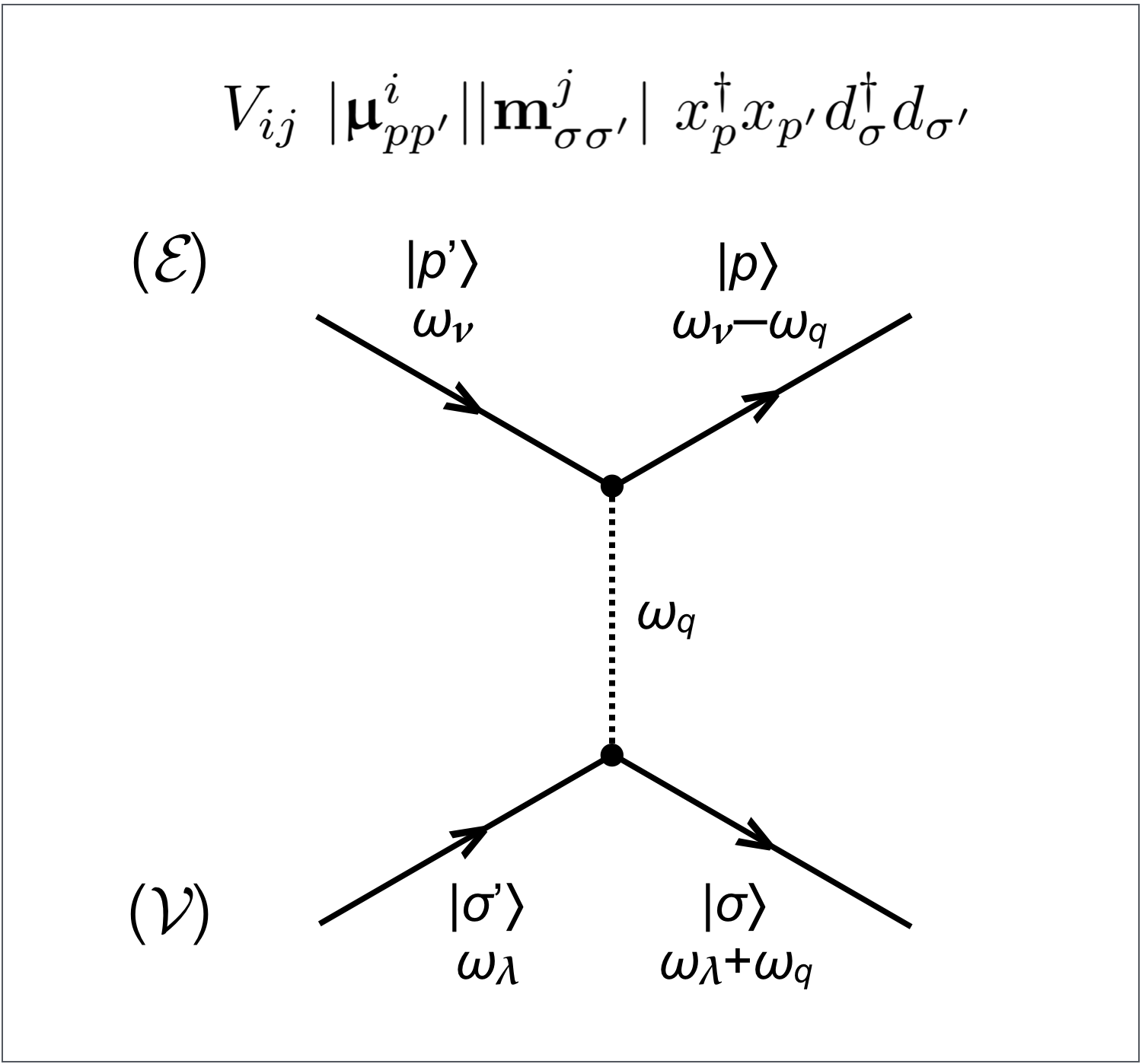}
\caption{\textbf{\textsf{Dipole-dipole interaction vertex.}} Diagrammatic representation of the dipole-dipole interaction between subsystems $(\mathcal{E})$ and $(\mathcal{V})$, as described by equation (\ref{dipdip}). This consists in a energy transfer mediated by a virtual photon of Matsubara frequency $\imath\omega_q$, while $\imath\omega_\nu$ and $\imath\omega_\lambda$ depict the Matsubara frequencies of the initial excited states of $(\mathcal{E})$ and $(\mathcal{V})$, respectively.}\label{vertex2}
\end{figure}

\vskip3mm
\noindent\textbf{Dipole-dipole interaction between $\bm{(\mathcal{E})}$ and $\bm{(\mathcal{V})}$.} The diagrammatic connection between subsystems $(\mathcal{E})$ and $(\mathcal{V})$ requires the knowledge of their interaction hamiltonian. In our case, QDs and ligands are expected to interact through dipolar coupling \cite{Noblet2018}. This is why we detail hereafter the quantum description of such an interplay between electric dipoles. 

Starting from the fermion operators defined above, the second quantization formalism enables us to expand the respective dipole moments of $(\mathcal{E})$ and $(\mathcal{V})$:
\begin{equation}\label{dipQD}
\bm{\upmu} = \sum_{p,p'} \left( \bm{\upmu}_{pp'}\ x_p^\dagger x_{p'} + \bm{\upmu}_{p'p}\ x_{p'}^\dagger x_{p} \right),
\end{equation}
\begin{equation}\label{dipMol}
\text{and} \hskip5mm \mathbf{m} = \sum_{\sigma,\sigma'} \left( \mathbf{m}_{\sigma\sigma'}\ d_\sigma^\dagger d_{\sigma'} + \mathbf{m}_{\sigma'\sigma}\ d_{\sigma'}^\dagger d_\sigma \right).
\end{equation}
The dipole-dipole interaction between the two subsystems is then described by:
\begin{equation}
\mathcal{H}_\text{dip-dip} = \bm{\upmu}^\text{T} \bm{V} \mathbf{m} = V_{ij} |\bm{\upmu}^i| |\mathbf{m}^j|,
\end{equation}
where $\bm{V}=(V_{ij})$ is the dipolar interaction matrix between the two dipole moments, given in Ref. \cite{Craig}. In view of equations (\ref{dipQD}) and (\ref{dipMol}), the terms of the interaction hamiltonian have the following form:
\begin{equation} \label{dipdip}
V_{ij}\ |\bm{\upmu}_{pp'}^i| |\mathbf{m}_{\sigma\sigma'}^j|\ x_p^\dagger x_{p'}d_\sigma^\dagger d_{\sigma'},
\end{equation}
and thus involve four-particle interactions. In terms of Feynman diagrams, such an interaction is depicted by the four-particle vertex which is presented in Figure \ref{vertex2}. This dipole-dipole vertex is the crux issue. It is the building block which enables us to formally couple $(\mathcal{E})$ and $(\mathcal{V})$, that is QDs and molecules, throughout the analytical derivation of the second-order susceptibility $\bm{\chi}^{(2)}$.

\begin{figure*}
\includegraphics[width=8.2cm]{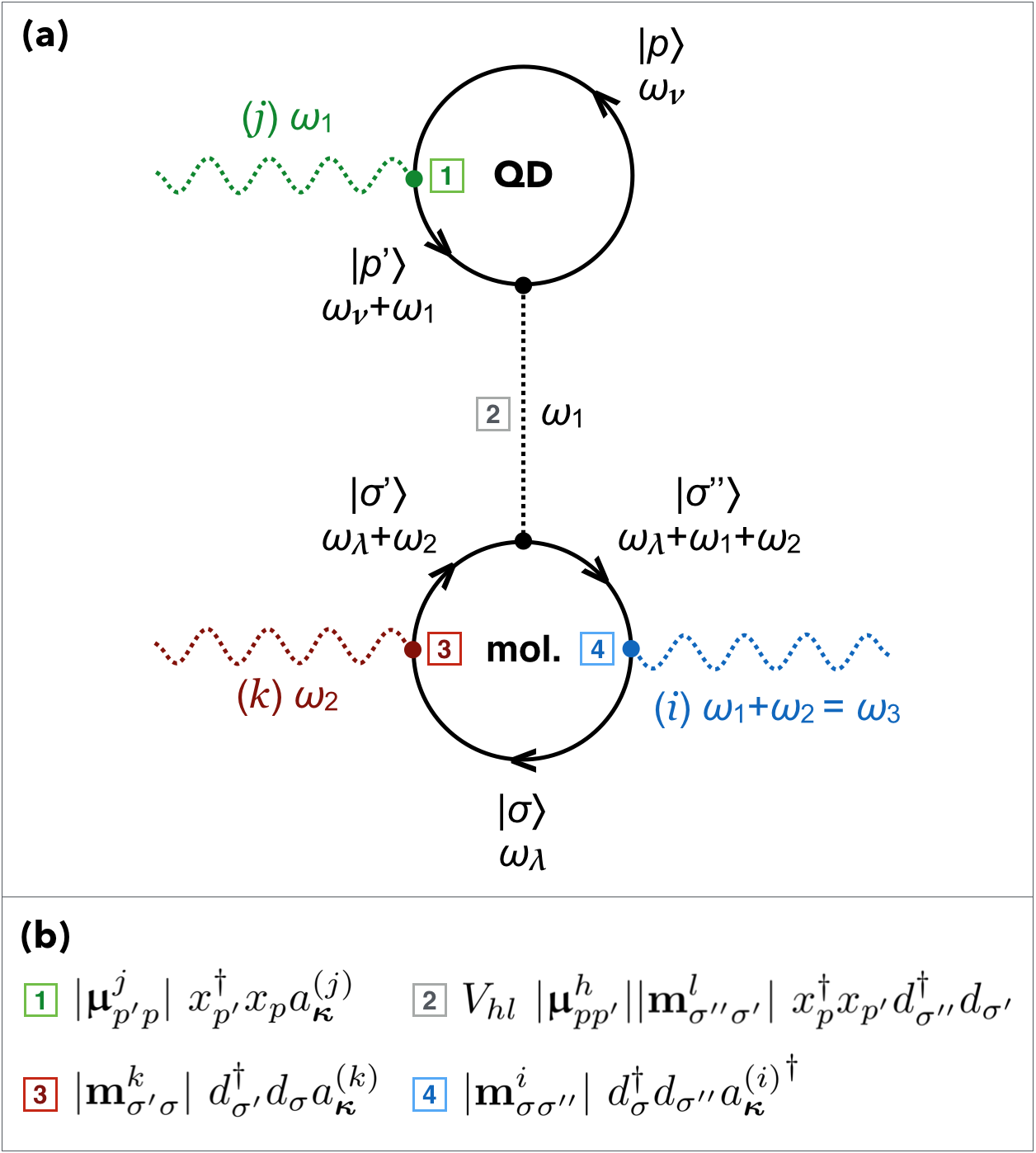} \hskip5mm \includegraphics[width=8.2cm]{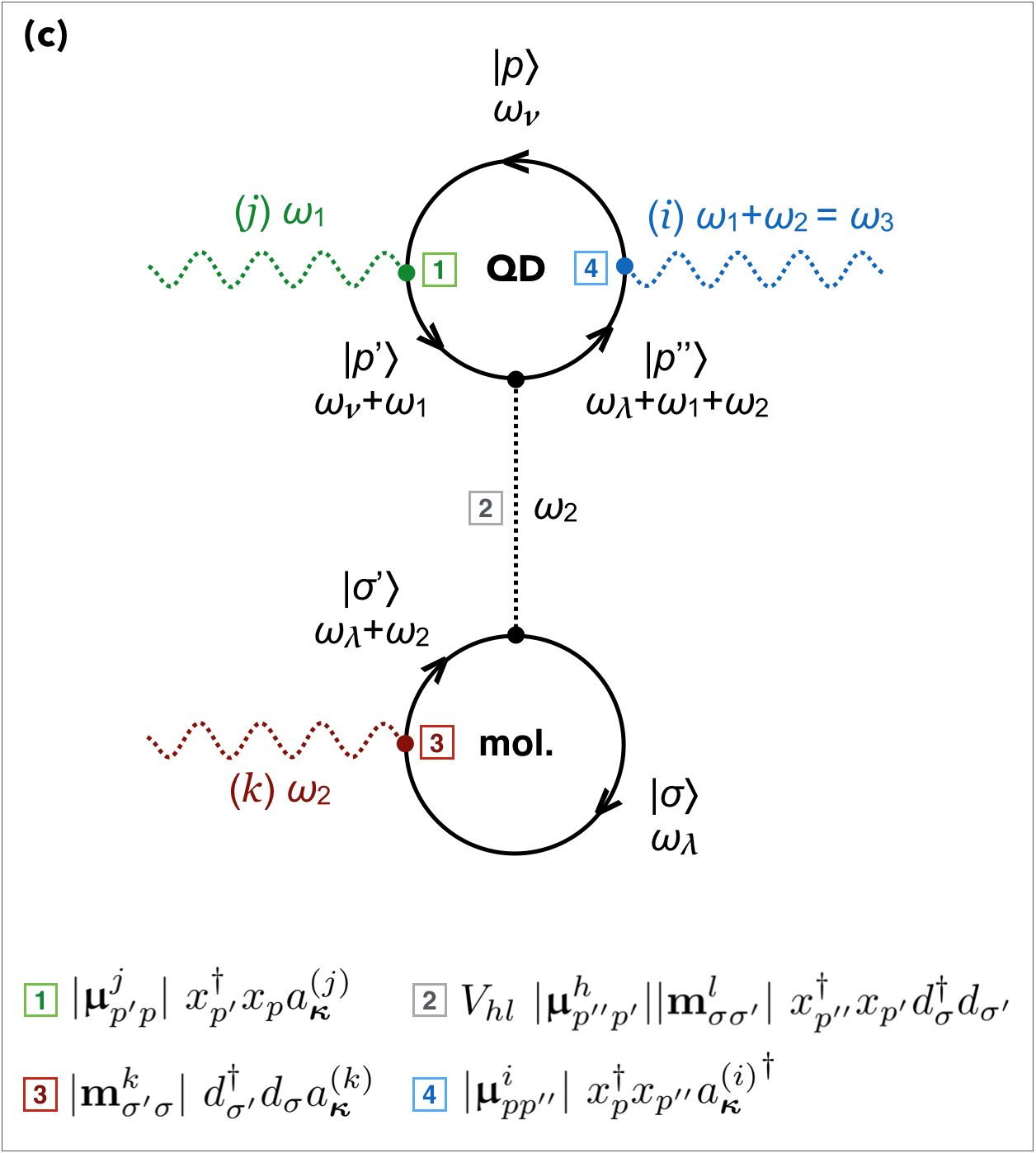}
\caption{\textbf{\textsf{Nonlinear response of hybrid systems. a}}, Diagrammatic representation of $\chi_{ijk}^{(2)}(\omega_1,\omega_2)$ accounting for the SFG process occurring in a QD/ligand complex. Each oriented line corresponds to the imaginary time Green's function associated to the propagation of a given state: $|p\rangle$ and $|p'\rangle$ depict the eigenstates of QDs, while $|\sigma\rangle$, $|\sigma'\rangle$ and $|\sigma''\rangle$ depict the eigenstates of ligands. \textbf{\textsf{b}}, Interaction terms associated to the four vertices with which the Feynman diagram of $\chi_{ijk}^{(2)}(\omega_1,\omega_2)$ is built. Especially, the scalar factors preceding the creation/annihilation operators explicitly intervene in the analytical derivation of the second-order susceptibility. The numbering refers to Fig. \ref{sfg}a.  \textbf{\textsf{c}}, Alternative diagrammatic representation of $\chi_{ijk}^{(2)}(\omega_1,\omega_2)$. As discussed in the article and Supplementary Note 3, this does not account for the molecular SFG response we are interested in.}\label{sfg}
\end{figure*}

\vskip3mm
\noindent\textbf{Diagrammatic calculation of $\bm{\chi^{(2)}}$.} Given the dipole-dipole and light-matter vertices featured in Figures \ref{vertex1} \& \ref{vertex2}, we propose to describe the nonlinear SFG coupling between QDs and molecules with the Feynman diagram drawn in Figure \ref{sfg}a. Within this picture, $\omega_1$ and $\omega_2$ respectively play the roles of the visible and infrared frequencies $\omega_\text{vis}$ and $\omega_\text{ir}$, as displayed in Figure \ref{principle}f. Since we are interested in the influence of QDs on the SFG signal of molecules, \emph{i.e.} the molecular SFG response, the Feynman diagram is built in such a way that the sum-frequency $\omega_3=\omega_1+\omega_2$ is generated by the molecule. First, as presented in Figure \ref{sfg}a \& b, the QD is excited by a $j$-polarized photon of frequency $\omega_1$ (Fig. \ref{sfg}b, vertex 1) and the molecule is excited by a $k$-polarized photon of frequency $\omega_2$ (Fig. \ref{sfg}b, vertex 3). Second, these excited states interact through dipole-dipole coupling (Fig. \ref{sfg}b, vertex 2), which translates into the exchange of a virtual photon of frequency $\omega_1$. Third and last, the molecule relaxes and emits an $i$-polarized photon of frequency $\omega_3$ (Fig. \ref{sfg}b, vertex 4). As a result, this Feynman diagram shows how a visible excitation (polarized along $j$) of QDs, combined to an infrared excitation (polarized along $k$) of molecules, leads to the molecular SFG response (polarized along $i$) of the hybrid system. Intuitively, this is thus expected to give access to the Green's function $\chi_{ijk}^{(2)}(\omega_1,\omega_2)$.

Naturally, a second diagram can be proposed, in which the SFG signal is generated by the QD. This possibility is presented in Figure \ref{sfg}c. As shown in Supplementary Note 3, this Feynman diagram results in a non-zero susceptibility only if the QD is non-centrosymmetric itself, which is a restrictive condition that is not necessary for the first diagram. Besides, as discussed further in this section, the expansion of $\chi_{ijk}^{(2)}(\omega_1,\omega_2)$ from the first diagram leads to a factorized form which involves the Raman and the infrared efficiencies of the molecule. This classical result, obtained so far in the conventional case of purely molecular systems, cannot arise from the diagram of Figure \ref{sfg}c (see Supplementary Note 3). Hence, the molecular SFG response is well and truly modeled by the Feynman diagram of Figure \ref{sfg}a.

As a second-order Green's function, $\chi_{ijk}^{(2)}$ is supposed to be expanded on the base of first-order Green's functions \cite{Doniach,Zagoskin,Kadanoff,Mahan}. This is the true meaning of Feynman diagrams, which actually represent perturbative expansions. As a consequence of the Feynman rules which drive such an expansion \cite{Zagoskin,Mahan}, the diagram of Figure \ref{sfg}a results in:
\begin{widetext}
\begin{eqnarray}
\chi^{(2)}_{ijk}(\imath\omega_1,\imath\omega_2) & = & \frac{1}{\beta} \sum_{p,p',\nu}  |\bm{\upmu}_{p'p}^j| |\bm{\upmu}_{pp'}^h|\ \mathfrak{G}_p(\imath\omega_\nu) \mathfrak{G}_{p'}(\imath\omega_\nu+\imath\omega_1) \label{sum1}\\
& & \times\ V_{hl} \nonumber\\
& & \times\ \frac{1}{\beta} \sum_{\sigma,\sigma',\sigma'',\lambda} |\mathbf{m}_{\sigma'\sigma}^k| |\mathbf{m}_{\sigma\sigma''}^i| |\mathbf{m}_{\sigma''\sigma'}^l|\ \mathfrak{G}_{\sigma}(\imath\omega_\lambda) \mathfrak{G}_{\sigma'}(\imath\omega_\lambda+\imath\omega_2)\mathfrak{G}_{\sigma''}(\imath\omega_\lambda+\imath\omega_3) \label{sum2}.
\end{eqnarray}
\end{widetext}
To simplify the notations in equation (\ref{sum1}-\ref{sum2}), the imaginary time Green's functions are itemized by a unique integer: $\mathfrak{G}_n(\imath\omega_\nu) = \mathfrak{G}_{nn}(\imath\omega_\nu) = (\imath\omega_\nu-\omega_n)^{-1}$, from equation (\ref{matsuGreen's}), in Methods. In virtue of equation (\ref{x1}), the sum over $p$, $p'$ and $\nu$ is identified to the linear susceptibility $\xi_{hj}=[\bm{\chi}^{(1)}_{\text{QD}}]_{hj}$ of QDs at frequency $\omega_1$. In equation (\ref{sum2}), the sum over $\lambda$ can be reduced according to Matsubara's theorem \cite{Zagoskin,Mahan}, reminded in Supplementary Note 1, and the sum over $\sigma$, $\sigma'$ and $\sigma''$ can be simplified by following these four assumptions: (i) the energy of the molecular ground state is set to zero ; (ii) the molecules remain in their ground state $|\sigma=0\rangle$ at equilibrium, \emph{i.e.} $\rho(\omega_\sigma)=\delta_{\sigma,0}$ ; (iii) the molecular states $|\sigma\rangle$ can be split into electronic and vibrational states, itemized by $|\sigma=e\rangle$ and $|\sigma=v\rangle$ ; (iv) the molecular SFG response is dominated by vibrational resonances with respect to the infrared frequency so that we only keep the terms with $(\imath\omega_2-\omega_v)^{-1}$. Under these hypotheses, we obtain:
\begin{equation}
\chi^{(2)}_{ijk}(\imath\omega_1,\imath\omega_2) = - \xi_{hj}(\imath\omega_1)\ V^h{}_{l} \sum_{e,v} \frac{|\mathbf{m}_{v0}^k| |\mathbf{m}_{0e}^i||\mathbf{m}_{ev}^l|}{(\imath\omega_2-\omega_v)(\imath\omega_3-\omega_e)},
\end{equation}
thanks to a calculation detailed in Supplementary Note 4. The sum over the molecular electronic states is known to give the molecular polarizability $\bm{\alpha}=(\alpha_{il})$ according to \cite{Lin1994}:
\begin{eqnarray}
\sum_e \frac{|\mathbf{m}_{0e}^i| |\mathbf{m}_{ev}^l|}{\imath\omega_3-\omega_e} & = & \langle 0| \sum_e \frac{|\mathbf{m}^i| |e\rangle \langle e| |\mathbf{m}^l|}{\imath\omega_3-\omega_e} |v\rangle \nonumber \\
& = &  \langle 0 | \alpha_{il}(\imath\omega_3) | v\rangle \nonumber \\
& = & \sqrt{\frac{\hbar}{2\omega_v}} \left.\frac{\partial\alpha_{il}}{\partial Q_v}\right|_0,
\end{eqnarray}
where $Q_v$ is the normal coordinate of vibration mode $|v\rangle$. For simplicity, we rewrite the derivative with respect to $Q_v$ as $\partial_v\alpha_{il}$. With the same notations, $|\mathbf{m}_{v0}^k| = \sqrt{\hbar/2\omega_v}\ \partial_vm_k$. Eventually, the second-order susceptibility reads:
\begin{equation}
\chi^{(2)}_{ijk}(\omega_1,\omega_2) = - \sum_{v} \frac{\hbar}{2\omega_v}\ \frac{\partial_v\alpha_{il}(\omega_3)\ V^{hl}\ \xi_{hj}(\omega_1)\ \partial_vm_k}{\omega_2-\omega_v+\imath 0^+},
\end{equation}
that is, in terms of tensors:
\begin{equation}\label{chi2}
\bm{\chi}^{(2)}(\omega_1,\omega_2) = - \sum_{v} \frac{\hbar}{2\omega_v}\ \frac{\partial_v\bm{\alpha}(\omega_3) \cdot \bm{V} \cdot \bm{\chi}_{\text{QD}}^{(1)}(\omega_1)\otimes \partial_v\mathbf{m}}{\omega_2-\omega_v+\imath 0^+}.
\end{equation}
For real systems, $\imath 0^+$ is replaced by the damping constant $\imath \gamma_v$ associated to vibration mode $|v\rangle$.

Here, the coupling between QDs and molecules explicitly appears through matrix $\bm{V}$, which drives their mutual dipolar interaction. In particular, this matrix couples the Raman response of molecules, via $\partial_v\bm{\alpha}(\omega_3)$, to the excitonic response of QDs, via $\bm{\xi}(\omega_1)=\bm{\chi}^{(1)}_{\text{QD}}(\omega_1)$. The infrared cross section  also intervenes with $\partial_v\mathbf{m}$. In respect of molecules, we notice that the SFG process combines their Raman and infrared cross sections. As announced above, we thus retrieve the classical result obtained in the case of purely molecular systems \cite{Lin1994}. For the hybrid system we are interested in, an additional contribution associated to QDs modulates the SFG process, whether these QDs are centrosymmetric or not, that is in spite of their ability to generate SFG on their own. The amplitude of each vibration mode $|v\rangle$ is driven by $\partial_v\bm{\alpha}(\omega_3) \cdot \bm{V} \cdot \bm{\chi}_{\text{QD}}^{(1)}(\omega_1)\otimes \partial_v\mathbf{m}$ and directly depends on the linear response function of QDs in the visible range.

\vskip7mm
\noindent \textbf{\textsf{Discussion}}

\vskip3mm \noindent We experimentally evidenced this coupling between QDs and molecules in the work reported in Ref. \cite{Noblet2018}. Within this previous article, a classical model based on dielectrics was proposed. It consists in assuming that the molecular dipole moment $\mathbf{m}$ experiences the local electric field $\mathbf{E}_\ell$ generated by the QD:
\begin{equation}\label{dip1}
\mathbf{m} = \bm{\alpha} \mathbf{E}_\ell + \mathbf{m}_0,
\end{equation}
where $\mathbf{m}_0$ is the dipole moment in absence of QD exciton. The local electric field $\mathbf{E}_\ell$ is related to the dipole moment $\bm{\upmu}$ of the QD, itself related to the excitation field $\mathbf{E}$ of the incident light through its linear susceptibility $\bm{\xi}=\bm{\chi}^{(1)}_{\text{QD}}$:
\begin{equation}\label{dip2}
\mathbf{E}_\ell = \bm{\kappa} \bm{\upmu} =  \bm{\kappa} \bm{\xi} \mathbf{E}.
\end{equation}
The matrix $\bm{\kappa}$ contains the spatial dependence of $\mathbf{E}_\ell$. For instance, in the static regime:
\begin{equation}
\kappa_{ij}(r)= \frac{3r_ir_j-\delta_{ij}r^2}{4\pi\varepsilon_0 r^5} \sim \frac{1}{r^3},
\end{equation}
wherein the origin is set at the center of the QD. Thanks to Fourier analysis, this model therefore results in \cite{Noblet2018}:
\begin{equation}\label{classical}
\bm{\chi}^{(2)}_v(\omega_1,\omega_2) \propto \frac{\partial_v\bm{\alpha}({\omega_1}) \cdot {\bm{\kappa}} \cdot \bm{\chi}_{\text{QD}}^{(1)}(\omega_1)\otimes {\mathbf{g}}}{\omega_2-\omega_v+\imath 0^+},
\end{equation}
for each vibration mode $|v\rangle$. We notice three differences with equation (\ref{chi2}): the $\omega_1$-dependence of $\partial_v\bm{\alpha}$, the matrix $\bm{\kappa}$ and the vector $\mathbf{g}$. It turns out that matrix $\bm{\kappa}$ plays the role of the dipolar interaction matrix $\bm{V}$, which is physically logical. Indeed, $\bm{\kappa}$ characterizes the dipolar interplay between the QD of moment $\bm{\upmu}$ and its ligands of moment $\mathbf{m}$ (see equations (\ref{dip1}-\ref{dip2})). In the static regime, $\bm{V}$ and $-\bm{\kappa}$ actually coincide \cite{Craig}:
\begin{equation}
V_{ij}(r)= \frac{\delta_{ij}r^2-3r_ir_j}{4\pi\varepsilon_0 r^5}.
\end{equation}
Here, the quantum and the classical approaches are equivalent. Nevertheless, the latter suffers from inaccuracies with respect to the Raman and the infrared cross sections. First, the classical derivation involves a Raman factor $\partial_v\bm{\alpha}$ at frequency $\omega_1$, whereas it is expected at frequency $\omega_3=\omega_1+\omega_2$ \cite{Lin1994}. As shown in equation (\ref{chi2}), the quantum and diagrammatic method allows the correction of this first inaccuracy. Second, the infrared efficiency $\partial_v\mathbf{m}$ does not appear in equation (\ref{classical}), contrary to equation (\ref{chi2}). Instead, the classical derivation requires the introduction of a vector $\mathbf{g}$ related to the geometry of the molecules which is not linked to $\partial_v\mathbf{m}$. More specifically, $\mathbf{g}$ depends on the mass-weighted cartesian coordinates of the atoms. Its precise definition is given in Ref. \cite{Noblet2018}. Consequently, the diagrammatic method we developed in the present article constitutes the proper way to compute second-order response functions for hybrid systems, and to resolve the deficiencies of classical approaches.

We chose to illustrate the diagrammatic method with SFG spectroscopy in the case of hybrid organic/inorganic systems because the understanding of the vibroelectronic coupling occurring within such systems is a controversial issue \cite{Humbert2015,Noblet2018,Swenson2016,Frederick2013,Lifshitz2015}. Given that our approach is very fundamental and general, this paves the way for the theoretical study of SHG (second harmonic generation), DFG (difference-frequency generation) and Raman spectroscopies. SHG and DFG are second-order processes, while Raman scattering is a third-order one. For a given system, the computation of SHG and DFG response functions would consist in drawing Feynman diagrams to connect three photons, in a way similar to what is done here for SFG. Besides, SHG is a particular case of SFG, for which $\omega_1=\omega_2$. Formally, the reasoning we used for the calculation of the SFG response tensor $\bm{\chi}^{(2)}(\omega_1,\omega_2)$ could thus be applied to the other second-order optical processes.

As a third-order process, Raman scattering is quite more complex to account for on a theoretical point of view \cite{Polavarapu1990}. Especially, when it is performed on molecules deposited on a rough metallic surface, this scattering is enhanced. One talks about surface-enhanced Raman spectroscopy (SERS), whose explanation is still a matter of debate \cite{Gersten1980,Moskovits1985,Otto1992,Stiles2008}. In this context, the present work may pave a new way of thinking for Raman and SERS with the difference that the associated Feynman diagrams would consist in combining four photons to compute $\bm{\chi}^{(3)}$ tensors. This may seem a daunting task, but we must keep in mind that the efficiency of the diagrammatic approach lies in mixing solid state and molecular physics with optics, through a unique formalism. For SERS, the question is precisely on the coupling between the metallic surface and the grafted molecules. This is why it is almost certain that our method will also yield positive results for organic/inorganic systems probed by third-order optical spectroscopies.

Fundamentally, Green's functions are the proper and dedicated language of linear and nonlinear responses theories. Although they may seem tricky to use for complex systems, this difficulty can be overcome thanks to Feynman diagrams. In the case of hybrid systems, our method consists in building such diagrams taking into account the interactions between the different subsystems, and to apply the Feynman rules which enables to translate any diagram into the analytical expression of the associated Green's function. Here, the case of ligand-conjugated QDs probed by SFG spectroscopy is considered to illustrate the method. Assuming a dipole-dipole interaction between QDs and ligands, the molecular SFG response proves to be modulated by the linear susceptibility of QDs in the visible range. The SFG process is thus expected to be the most efficient when the visible excitation of QDs is maximum. As shown in Ref. \cite{Noblet2018}, such a behaviour has been observed. Our theoretical result therefore coincides with the experiment, which encourages the use of this approach in nonlinear optics to describe and understand the behaviour of such organic/inorganic hybrid systems.

\vskip7mm

\noindent\textbf{\textsf{Methods}}\small
\vskip3mm
\noindent\textbf{\textsf{Green's functions in optics and solid-state physics.}} Probing matter with light consists in examining how the polarization of matter, $\mathbf{P}$, reacts to an excitation electric field $\mathbf{E}$. In frequency space, the linear response theory teaches that there exists a tensor $\bm{\chi}^{(1)}=(\chi_{ij}^{(1)})$ so that \cite{Boyd}:
\begin{equation}
\mathbf{P}(\omega) = \bm{\chi}^{(1)}(\omega) \mathbf{E}(\omega).
\end{equation}
In time domain, this gives: $\mathbf{P}(t) = \bm{\chi}^{(1)} \ast \mathbf{E}(t)$. To go further, the use of an intense light allows studying the behaviour of matter beyond the linear dielectric regime. In this case, a nonlinear polarization $\mathbf{P}_\text{NL}$ is induced in the material which reveals the existence of a second-order response tensor $\bm{\chi}^{(2)}=(\chi_{ijk}^{(2)})$, so that \cite{Boyd}:
\begin{equation}
\mathbf{P}_\text{NL}(t) = \bm{\chi}^{(2)} \ast \mathbf{E}\otimes\mathbf{E} (t).
\end{equation} 
As response functions, each $\chi_{ij}^{(1)}$ and $\chi_{ijk}^{(2)}$ are Green's functions and thus belongs to a large function family which exceeds the scope of optics \cite{Greiner}. In condensed matter physics, Green's functions are used to describe the propagation of excited states in solids \cite{Doniach,Zagoskin,Kadanoff,Mahan}. More precisely, for a quantum system characterized by the eigenstates $|n\rangle$, $n\in\mathbb{N}$, and prepared in state $|m\rangle$ at $t=0$, the retarded Green's function $G_{mn}(t)$ gives the complex amplitude of probability to measure the system in state $|n\rangle$ at time $t$ \cite{Zagoskin,Mahan}. In particular, $G_{mm}(t)$ measures the coherence of each state $|m\rangle$ over time. Given those two subfamilies of Green's functions, our theoretical approach aims to connect both.

\vskip3mm
\noindent\textbf{\textsf{Feynman diagrams in Matsubara frequency space.}} For convenience, we choose to handle imaginary time Green's functions $\mathfrak{G}_{nm}$ instead of $G_{nm}$ \cite{Zagoskin,Mahan}. In frequency domain, both are related through analytical continuity and then constitute two equivalent mathematical pictures \cite{Mahan}:
\begin{equation}
G_{nm}(\omega) = \mathfrak{G}_{nm}(\omega+\imath0^+).
\end{equation}
Given the eigenfrequencies $\omega_n$ associated to the eigenstates $|n\rangle$, the imaginary time Green's functions prove to simply read \cite{Zagoskin,Mahan}:
\begin{equation}\label{matsuGreen's}
\mathfrak{G}_{nm}(\imath\omega_\nu) = \frac{\delta_{nm}}{\imath\omega_\nu - \omega_n},
\end{equation}
where $\imath\omega_\nu$, $\nu\in\mathbb{Z}$, are the Matsubara frequencies, that is the poles of the Fermi-Dirac function $\rho(z)=(1+e^{\hbar\beta z})^{-1}$, $z\in\mathbb{C}$ (see Supplementary Note 1). Within the formalism of Feynman diagrams, each $\mathfrak{G}_{nm}(\imath\omega_\nu)$ can be represented as an oriented straight line embodying the propagation of $|n\rangle$ with energy $\omega_\nu$ \cite{Zagoskin,Mahan}. When different lines are crossing at a vertex, it means that several states interact with each other. Thereby, drawing a Feynman diagram consists in representing the propagation of an excited state according to the interactions that it experiences. In virtue of Feynman rules \cite{Zagoskin,Mahan}, the Green's function which drives the whole process is computed as a combination of all the Green's functions associated to the straight lines which compose the whole diagram.

\vskip3mm
\noindent\textbf{\textsf{Light-matter interaction.}} To draw Feynman diagrams accounting for optical response functions, we need the unit diagrams associated to the interaction between the quantum states of matter and light. As we know, light is described by a quantized electric field \cite{Craig}:
\begin{equation}
\mathbf{E} = \imath \sum_{\bm{\kappa},i} \sqrt{\frac{\hbar\omega}{2V\varepsilon_0}} \left( \textbf{e}^{(i)} a_{\bm{\kappa}}^{(i)} - \bar{\textbf{e}}^{(i)} {a_{\bm{\kappa}}^{(i)}}^\dagger \right),
\end{equation}
where $a_{\bm{\kappa}}^{(i)}$ and ${a_{\bm{\kappa}}^{(i)}}^\dagger$ are boson operators associated to $i$-polarized photons of momentum $\bm{\kappa}$. Therefore, the interaction between light and any quantum system described by equation (\ref{eqG}) translates into:
\begin{equation}\label{lightmat}
\mathcal{H}_{\text{light-mat.}} = -\mathbf{p} \cdot \mathbf{E},
\end{equation}
wherein $\mathbf{p}$ is the dipole moment of the system, whose second quantization expansion reads:
\begin{equation}\label{dip}
\mathbf{p} = \sum_{n,m} \left( \mathbf{p}_{nm}\ c_n^\dagger c_{m} + \mathbf{p}_{mn}\ c_{m}^\dagger c_{n} \right).
\end{equation}
Equation (\ref{lightmat}) then involves terms of the forms: 
\begin{equation}\label{abs}
|\mathbf{p}_{nm}^i|\ c_n^\dagger c_m a_{\bm{\kappa}}^{(i)}\ \text{and}\ |\mathbf{p}_{mn}^i|\ c_m^\dagger c_n{a_{\bm{\kappa}}^{(i)}}^\dagger.
\end{equation}
The associated three-particle vertices are presented in Figure \ref{vertex1}a \& b. These embody light absorption and light emission processes.

\bibliography{biblio}

\vskip7mm
\noindent \textbf{\textsf{Acknowledgements}} \small

\noindent The authors do want to thank Dr B. Busson and Dr. A. Tadjeddine (University of Paris-Sud, Universit\'e Paris-Saclay, Laboratoire de Chimie Physique, CNRS, France), and Dr F. Hache (Laboratoire d'Optique et Biosciences, CNRS-INSERM, Ecole Polytechnique, France) for fruitful scientific discussion.

\vskip7mm
\noindent \textbf{\textsf{Author contributions}} \small

\noindent The manuscript was achieved through contributions of all authors. T.N. developed the theoretical approach based on imaginary time Green's functions and Feynman diagrams, drew all the diagrams and derived the corresponding analytical expressions of the optical response functions. T.N. and C.H. led and realized the entire writing process.

\end{document}


\noindent \Large\textbf{{Sum-frequency generation through a unique Feynman diagram formalism: the case of bipartite organic/inorganic complexes}}
\normalsize
\vskip1cm
\noindent T. Noblet $\cdot$ C. Humbert

\vskip3cm

\section*{Supplementary Note 1}

\noindent \textsf{\textbf{Matsubara frequencies and Matsubara's theorem.}} The Matsubara frequencies $\{\imath\omega_\nu\}_{\nu\in\mathbb{Z}}$ of fermions are defined as the poles of the Fermi-Dirac functions $\rho(z)=(e^{\hbar\beta z}+1)^{-1}$:
\begin{equation}
\hbar\omega_\nu = \frac{\pi}{\beta}\ (2\nu+1).
\end{equation}
For bosons, these are the poles of the Bose-Einstein function $\varrho(z)=(e^{\hbar\beta z}-1)^{-1}$:
\begin{equation}
\hbar\omega_\gamma = \frac{\pi}{\beta}\ 2\gamma,\ \gamma\in\mathbb{Z}.
\end{equation}
In this way, for bosons: $e^{\imath\hbar\beta \omega_\gamma}=1$. It means that $\rho(\omega\pm\imath\omega_\gamma)=\rho(\omega)$. Since photons are bosons, this identity is used to simplify the computation of optical response functions.

Besides, Matsubara's theorem allows the reduction of sums of products of imaginary time Green's functions. It states that for all meromorphic function $\phi$ characterized by $N$ poles $\{z_u\}_{1\leqslant u\leqslant N}$ and the associated residues $\{r_u\}_{1\leqslant u\leqslant N}$ \cite{Zagoskin,Mahan}:
\begin{equation}\label{matsu}
\frac{1}{\beta} \sum_{\nu} \phi(\imath\omega_\nu) = \sum_u r_u\ \rho(z_u).
\end{equation}
In the present article, we deal with two kinds of meromorphic functions:
\begin{equation}
\phi_1(z) = \frac{1}{z-a}\ \frac{1}{z-b} \hskip5mm \text{and} \hskip5mm \phi_2(z) = \frac{1}{z-a}\ \frac{1}{z-b}\ \frac{1}{z-c},
\end{equation}
wherein $a$, $b$ and $c$ denote the poles.
The function $\phi_1$ is involved in the computation of linear response functions, and $\phi_2$ in the computation of second-order response functions. From equation (\ref{matsu}), we get:
\begin{equation}\label{matsu1}
\frac{1}{\beta} \sum_{\nu} \phi_1(\imath\omega_\nu) = \frac{\rho(a)-\rho(b)}{a-b},
\end{equation}
and:
\begin{equation}\label{matsu2}
\frac{1}{\beta} \sum_{\nu} \phi_2(\imath\omega_\nu) = \frac{\rho(a)}{(a-b)(a-c)} + \frac{\rho(b)}{(b-a)(b-c)} + \frac{\rho(c)}{(c-a)(c-b)}.
\end{equation}

\vskip1cm
\section*{Supplementary Note 2}

\noindent \textsf{\textbf{Diagrammatic method applied to doubly resonant molecular systems.}} If the diagrammatic method proposed in the article is applied to hybrid systems, it is not dedicated to this only case. It is also interesting for the conventional case of purely molecular systems. Here, we examine the second-order susceptibility associated to the Feynman diagram given in Supplementary Fig. 1a.

Applying the Feynman rules, we obtain:
\begin{equation}\label{eq1}
\chi_{ijk}^{(2)}(\imath\omega_1,\imath\omega_2) = \frac{1}{\beta} \sum_{\sigma,\sigma',\sigma'',\lambda} m^i_{\sigma\sigma''} m^j_{\sigma'\sigma} m^k_{\sigma''\sigma'}\ \mathfrak{G}_\sigma(\imath\omega_\lambda) \mathfrak{G}_{\sigma'}(\imath\omega_\lambda+\imath\omega_1) \mathfrak{G}_{\sigma''}(\imath\omega_\lambda+\imath\omega_3).
\end{equation}
Thanks to Matsubara's theorem (see Supplementary Note 1), we reduce the sum over $\lambda$:
\begin{eqnarray}
& & \frac{1}{\beta} \sum_{\lambda} \mathfrak{G}_\sigma(\imath\omega_\lambda) \mathfrak{G}_{\sigma'}(\imath\omega_\lambda+\imath\omega_1) \mathfrak{G}_{\sigma''}(\imath\omega_\lambda+\imath\omega_3)\nonumber\\
& = & \frac{1}{\beta} \sum_{\lambda} \frac{1}{\imath\omega_\lambda -\omega_\sigma}\ \frac{1}{\imath\omega_\lambda -(\omega_{\sigma'}-\imath\omega_1)}\ \frac{1}{\imath\omega_\lambda -(\omega_{\sigma''}-\imath\omega_3)}\nonumber\\
& = & \frac{\rho(\omega_\sigma)}{(\omega_\sigma-\omega_{\sigma'}+\imath\omega_1)(\omega_\sigma-\omega_{\sigma''}+\imath\omega_3)}\nonumber\\
& & +\ \frac{\rho(\omega_{\sigma'}-\imath\omega_1)}{(\omega_{\sigma'}-\omega_{\sigma}-\imath\omega_1)(\omega_{\sigma'}-\omega_{\sigma''}+\imath\omega_2)}\label{eq2}\\
& & +\ \frac{\rho(\omega_{\sigma''}-\imath\omega_3)}{(\omega_{\sigma''}-\omega_{\sigma}-\imath\omega_3)(\omega_{\sigma''}-\omega_{\sigma'}-\imath\omega_2)}.\nonumber
\end{eqnarray}
Since photons are bosons: $\rho(\omega_{\sigma'}-\imath\omega_1)=\rho(\omega_{\sigma'})$ and $\rho(\omega_{\sigma''}-\imath\omega_3)=\rho(\omega_{\sigma''})$. Assuming that the molecule is in its ground state $|0\rangle$ at equilibrium, \emph{i.e.} $\rho(\omega_\sigma)=\delta_{\sigma 0}$, setting the ground state energy to zero, and distinguishing the electronic and vibrational states, $|\sigma=e,v\rangle$, equations (\ref{eq1}-\ref{eq2}) give:
\begin{eqnarray}
\sum_{e,v} & \Bigg( & \frac{m_{0e}^i m_{v0}^jm_{ev}^k}{(\imath\omega_1-\omega_v)(\imath\omega_3-\omega_e)} + \frac{m_{0v}^i m_{e0}^jm_{ve}^k}{(\imath\omega_1-\omega_e)(\imath\omega_3-\omega_v)}\label{res2} \\
& - & \frac{m_{ev}^i m_{0e}^jm_{v0}^k}{(\imath\omega_1+\omega_e)(\imath\omega_2-\omega_v)} - \frac{m_{ve}^i m_{0v}^jm_{e0}^k}{(\imath\omega_1+\omega_v)(\imath\omega_2-\omega_e)}\label{res1}\\
& + & \frac{m_{e0}^i m_{ve}^jm_{0v}^k}{(\imath\omega_3+\omega_e)(\imath\omega_2+\omega_v)} +\frac{m_{v0}^i m_{ev}^jm_{0e}^k}{(\imath\omega_3+\omega_v)(\imath\omega_2+\omega_e)} \hskip3mm \Bigg).
\end{eqnarray}
Implicitly, we also assumed that there is not any optical transition between two different electronic states or two vibrational states. Indeed, the aim is to account for SFG processes described by the canonical energy diagram presented in Supplementary Fig. 1b.

\begin{figure}
\centering
\includegraphics[width=15cm]{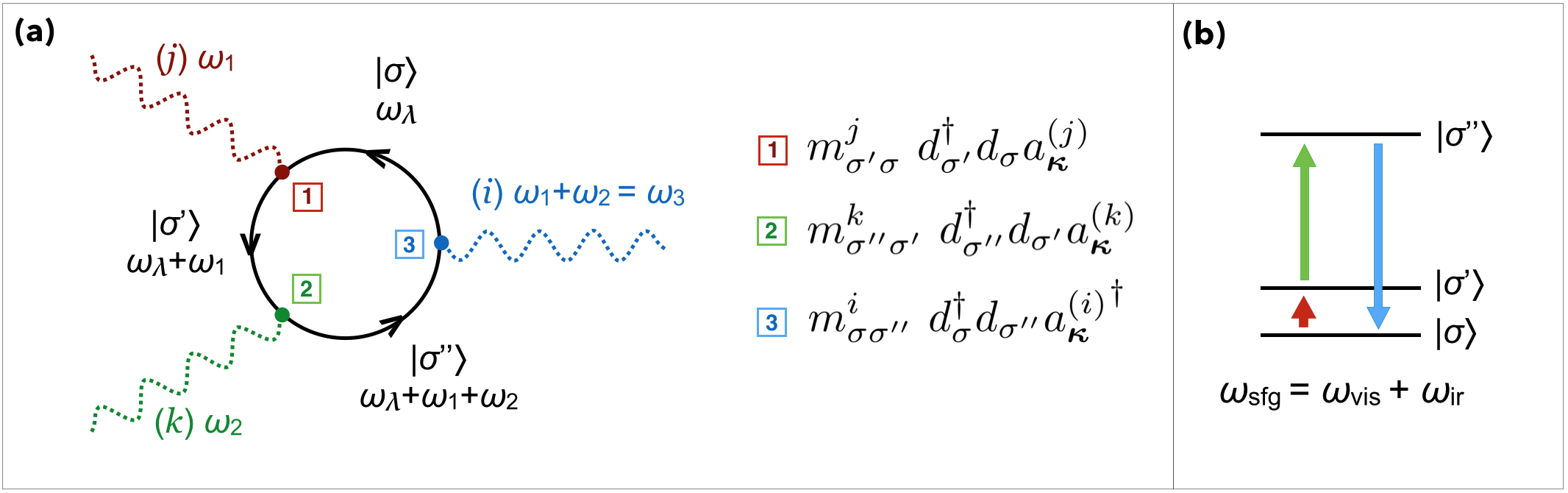}
\caption*{\textbf{Supplementary Figure 1} -- \textbf{\textsf{(a)}} Diagrammatic representation of $\chi^{(2)}_{ijk}(\imath\omega_1,\imath\omega_2)$ accounting for the SFG process in a purely molecular system. \textbf{\textsf{(b)}} Energy diagram illustrating the doubly resonant SFG process.}
\end{figure}

For such a purely molecular system, there are two cases to examine:\\
$\bullet$ \textbf{1st case:} $\omega_1=\omega_\text{vis}$ and $\omega_2=\omega_\text{ir}$. The first term of equation (\ref{res1}) is then the only one to exhibit a vibrational resonance with respect to the infrared light, \emph{via} $(\imath\omega_2-\omega_v)^{-1}$. However, this term is not resonant with respect to the visible light, because of $(\imath\omega_1+\omega_e)^{-1}$.\\
$\bullet$ \textbf{2nd case:} $\omega_1=\omega_\text{ir}$ and $\omega_2=\omega_\text{vis}$. The first term of equation (\ref{res2}) is thus the only adequate one, accounting for a double resonance thanks to $(\imath\omega_1-\omega_v)^{-1}$ and $(\imath\omega_3-\omega_e)^{-1}$.

Consequently, the Green's function associated to the doubly resonant SFG response is:
\begin{equation}
\chi_{ijk}^{(2)}(\omega_\text{ir},\omega_\text{vis}) = \sum_{v,e} \frac{m_{0e}^i m_{v0}^jm_{ev}^k}{(\omega_\text{ir}-\omega_v+\imath0^+)(\omega_\text{sfg}-\omega_e+\imath0^+)} 
\end{equation}
The sum over $e$ is known to gives the resonant part of the polarizability \cite{Lin1994}, assimilated to the entire polarizability:
\begin{equation}
\sum_{e} \frac{m_{0e}^i m_{ev}^k}{\omega_\text{sfg}-\omega_e+\imath0^+} = \langle 0| \alpha_{ik}(\omega_\text{sfg})| v \rangle = \sqrt{\frac{\hbar}{2\omega_v}}\ \left.\frac{\partial\alpha_{ik}}{\partial Q_v}\right|_0^{\omega_\text{sfg}}.
\end{equation}
Moreover:
\begin{equation}
m_{v0}^j = \langle v| m_{j} | 0 \rangle = \sqrt{\frac{\hbar}{2\omega_v}}\ \left.\frac{\partial m_{j}}{\partial Q_v}\right|_0.
\end{equation}
Eventually, we obtain:
\begin{equation}\label{chi2mol}
\chi_{ijk}^{(2)}(\omega_\text{ir},\omega_\text{vis}) = \sum_{v} \frac{\hbar}{2\omega_v}\ \left.\frac{\partial m_{j}}{\partial Q_v}\right|_0 \left.\frac{\partial\alpha_{ik}}{\partial Q_v}\right|_0^{\omega_\text{sfg}} \frac{1}{\omega_\text{ir}-\omega_v+\imath0^+}.
\end{equation}
We actually retrieve the expected formula for doubly resonant molecular systems \cite{Hayashi2002,Hung2015}, which proves the relevance and the accuracy of our diagrammatic method.

Strictly speaking, equation (\ref{chi2mol}) does not correspond to the susceptibility but to the molecular hyperpolarizability $\beta_{ijk}$ (not linked to the notation $\beta=(k_BT)^{-1}$ used in the article). The former is a macroscopic quantity characterizing the whole molecular system, while the latter contains the microscopic information about a single molecule. In a proper way, we shall write:
\begin{equation}
\chi_{ijk}^{(2)} = N \langle \beta_{ijk} \rangle_\text{mol},
\end{equation}
where $N$ is the density of molecules and $\langle \cdot \rangle_\text{mol}$ the average over all the molecules.

\vskip1cm
\section*{Supplementary Note 3}

\noindent \textsf{\textbf{Derivation of the Green's function associated to Figure 4c.}} As mentioned in the article, the Feynman diagram presented in Figure 4c does not account for the molecular SFG response we are interested in for the hybrid QD/ligands system. Applying Feynman rules, we obtain:
\begin{eqnarray}
\chi^{(2)}_{ijk}(\imath\omega_1,\imath\omega_2) & = & \frac{1}{\beta} \sum_{p,p',p'',\nu}  \mu_{p'p}^j\mu_{pp''}^i\mu_{p''p'}^h\ \mathfrak{G}_p(\imath\omega_\nu) \mathfrak{G}_{p'}(\imath\omega_\nu+\imath\omega_1) \mathfrak{G}_{p''}(\imath\omega_\nu+\imath\omega_3) \label{sum1}\\
& & \times\ V_{hl} \nonumber\\
& & \times\ \frac{1}{\beta} \sum_{\sigma,\sigma',\lambda} m_{\sigma'\sigma}^km_{\sigma\sigma'}^l\ \mathfrak{G}_{\sigma}(\imath\omega_\lambda) \mathfrak{G}_{\sigma'}(\imath\omega_\lambda+\imath\omega_2) \label{sum2}.
\end{eqnarray}
Given Supplementary Note 2, we easily understand that equation (\ref{sum1}) translates into the hyperpolarizability $\beta_{ijh}^\text{QD}(\omega_1,\omega_2)$ of a QD and that equation (\ref{sum2}) results in the molecular polarizability $\alpha_{lk}(\omega_2)$. Therefore:
\begin{equation}
\bm{\chi}^{(2)}(\omega_1,\omega_2) =- \bm{\beta}^\text{QD}(\omega_1,\omega_2) \cdot \bm{V} \cdot \bm{\alpha}(\omega_2).
\end{equation}
First, the vibrational resonance with respect to $\omega_2=\omega_\text{ir}$ is hidden within the molecular polarizability $\bm{\alpha}(\omega_2)$, hence our inability to extract the Raman and infrared cross sections. Second, the nonlinear response of QDs is directly involved. As explained in the article, their hyperpolarizability $\bm{\beta}^\text{QD}$ is non-zero only if they are non-centrosymmetric, which constitues a restrictive and unnecessary condition.

To conclude, the Feynman diagram of Figure 4c is not proper to describe the impact of QDs on the molecular SFG response.

\vskip1cm

\section*{Supplementary Note 4}

\noindent \textsf{\textbf{Details about the derivation of the Green's function associated to Figure 4a.}} The article presents the derivation of $\bm{\chi}^{(2)}$ for a hybrid system made of ligand-conjugated QDs. A complete version is detailed here. Applying the Feynman rules to the diagram of Figure 4a, we obtain:
\begin{eqnarray}
\chi^{(2)}_{ijk}(\imath\omega_1,\imath\omega_2) & = & \frac{1}{\beta} \sum_{p,p',\nu}  \mu_{p'p}^j\mu_{pp'}^h\ \mathfrak{G}_p(\imath\omega_\nu) \mathfrak{G}_{p'}(\imath\omega_\nu+\imath\omega_1) \label{sumA}\\
& & \times\ V_{hl} \nonumber\\
& & \times\ \frac{1}{\beta} \sum_{\sigma,\sigma',\sigma'',\lambda} m_{\sigma'\sigma}^km_{\sigma\sigma''}^im_{\sigma''\sigma'}^l\ \mathfrak{G}_{\sigma}(\imath\omega_\lambda) \mathfrak{G}_{\sigma'}(\imath\omega_\lambda+\imath\omega_2)\mathfrak{G}_{\sigma''}(\imath\omega_\lambda+\imath\omega_3) \label{sumB}.
\end{eqnarray}
The sum (\ref{sumA}) is identified to the linear susceptibility of QDs \cite{Lin1994}, noted  $\xi_{hj}=[\bm{\chi}^{(1)}_{\text{QD}}]_{hj}$. From Matsubara's theorem (see Supplementary Note 1), we reduce the sum over $\lambda$:
\begin{eqnarray}
& & \displaystyle \frac{1}{\beta} \sum_{\lambda}  \mathfrak{G}_{\sigma}(\imath\omega_\lambda) \mathfrak{G}_{\sigma'}(\imath\omega_\lambda+\imath\omega_2)\mathfrak{G}_{\sigma''}(\imath\omega_\lambda+\imath\omega_3)\nonumber \\
& = & \frac{\rho(\omega_\sigma)}{(\omega_\sigma-\omega_{\sigma'}+\imath\omega_2)(\omega_\sigma - \omega_{\sigma''}+\imath\omega_3)}\nonumber\\
& & +\ \frac{\rho(\omega_{\sigma'}-\imath\omega_2)}{(\omega_{\sigma'}-\omega_\sigma-\imath\omega_2)(\omega_{\sigma'}-\omega_{\sigma''}+\imath\omega_1)}\label{eq}\\
& & +\ \frac{\rho(\omega_{\sigma''}-\imath\omega_3)}{(\omega_{\sigma''}-\omega_\sigma-\imath\omega_3)(\omega_{\sigma''}-\omega_{\sigma'}-\imath\omega_1)}.\nonumber
\end{eqnarray}
Since photons are bosons: $\rho(\omega_{\sigma'}-\imath\omega_2)=\rho(\omega_{\sigma'})$ and $\rho(\omega_{\sigma''}-\imath\omega_3)=\rho(\omega_{\sigma''})$. Assuming that the molecule is in its ground state $|0\rangle$ at equilibrium, \emph{i.e.} $\rho(\omega_\sigma)=\delta_{\sigma 0}$, setting the ground state energy to zero, and distinguishing the electronic and vibrational states, $|\sigma=e,v\rangle$, equations (\ref{sumB}-\ref{eq}) give:
\begin{eqnarray}
\sum_{v,e} &\bigg( & \frac{m_{e0}^km_{0v}^im_{ve}^l}{(\imath\omega_2-\omega_e)(\imath\omega_3- \omega_v)} + \frac{m_{v0}^km_{0e}^im_{ev}^l}{(\imath\omega_2-\omega_v)(\imath\omega_3- \omega_e)}\label{sumev}\\
 & - & \frac{m_{0e}^km_{ev}^im_{v0}^l}{(\imath\omega_2+\omega_e)(\imath\omega_1-\omega_v)} - \frac{m_{0v}^km_{ve}^im_{e0}^l}{(\imath\omega_2+\omega_v)(\imath\omega_1-\omega_e)}\\
 & + & \frac{m_{ev}^km_{v0}^im_{0e}^l}{(\imath\omega_3+\omega_v)(\imath\omega_1+\omega_e)} + \frac{m_{ve}^km_{e0}^im_{0v}^l}{(\imath\omega_3+\omega_e)(\imath\omega_1+\omega_v)} \hskip3mm \bigg).
\end{eqnarray}
For the same reason than that invoked in Supplementary Note 2, we also assumed that there is not any optical transition between two different electronic states or two vibrational states.

Since $\omega_1=\omega_\text{vis}$ and $\omega_2=\omega_\text{ir}$, the second term of equation (\ref{sumev}) is the only adequate one, accounting for a vibrational resonance with respect to the infrared light \emph{via} $(\imath\omega_2-\omega_v)^{-1}$. We thus obtain:
\begin{equation}
\chi^{(2)}_{ijk}(\imath\omega_1,\imath\omega_2) = - \xi_{hj}(\imath\omega_1)\ V^h{}_{l} \sum_{e,v} \frac{m_{v0}^km_{0e}^i m_{ev}^l}{(\imath\omega_2-\omega_v)(\imath\omega_3-\omega_e)}.
\end{equation}
For the same reason than that given in Supplementary Note 2, the sum over the molecular electronic states results in the molecular polarizability $\bm{\alpha}=(\alpha_{il})$:
\begin{equation}
\sum_e \frac{m_{0e}^i m_{ev}^l}{\imath\omega_3-\omega_e} =  \langle 0| \sum_e \frac{m^i|e\rangle \langle e| m^l}{\imath\omega_3-\omega_e} |v\rangle \nonumber  =  \langle 0 | \alpha_{il}(\imath\omega_3) | v\rangle = \sqrt{\frac{\hbar}{2\omega_v}} \left.\frac{\partial\alpha_{il}}{\partial Q_v}\right|_0^{\omega_3}.
\end{equation}
Since $m_{v0}^k= \langle v| m_{k} | 0 \rangle = \sqrt{\frac{\hbar}{2\omega_v}} \left.\frac{\partial m_{k}}{\partial Q_v}\right|_0$, we eventually get:
\begin{equation}
\chi_{ijk}^{(2)}(\omega_1,\omega_2) = -\sum_{v} \frac{\hbar}{2\omega_v}\ \left.\frac{\partial\alpha_{il}}{\partial Q_v}\right|_0^{\omega_3} \frac{V^{hl}\ \xi_{hj}(\omega_1)}{\omega_2-\omega_v+\imath0^+}\ \left.\frac{\partial m_{k}}{\partial Q_v}\right|_0,
\end{equation}
which indeed corresponds to the result given in the article.

\newpage

\section*{Supplementary References}

\bibliographystyle{naturemag}
\bibliography{bibliosup}